\pdfoutput=1
\documentclass[conference]{IEEEtran}
\ifCLASSINFOpdf
\else
   \usepackage{graphicx}
   \graphicspath{{../eps/}}
   \DeclareGraphicsExtensions{.eps}
\fi

\usepackage{amsmath}
\usepackage{amsfonts,amssymb}
\usepackage{amssymb}
\usepackage{wrapfig}
\usepackage{psfrag}
\usepackage{epstopdf}
\usepackage{cite}
\usepackage{graphicx}
\usepackage{subfigure}
\usepackage{threeparttable}
\usepackage{cases}
\usepackage{subeqnarray}
\usepackage{color}
\usepackage{underscore}
\usepackage{verbatim}
\usepackage{bm}
\usepackage{stfloats}

\newtheorem{theorem}{Theorem}
\newtheorem{lemma}{Lemma}
\newtheorem{corollary}{Corollary}

\newcommand{\sinc}{\mathrm{sinc}}
\usepackage{algorithm}
\usepackage{algorithmic}

\begin{document}

\title{Wireless Communication Using Metal Reflectors: Reflection Modelling and Experimental Verification}
%
%
%

\author{\IEEEauthorblockN{$\text{Zhi~Yu}^*$, $\text{Chao~Feng}^*$, $\text{Yong~Zeng}^*\ddagger$, $\text{Teng~Li}^\dagger\ddagger$, and $\text{Shi~Jin}^*$}
\IEEEauthorblockA{*National Mobile Communications Research Laboratory, Southeast University, Nanjing 210096, China\\
$\dagger$State Key Laboratory of Millimeter Waves, Southeast University, Nanjing 210096, China\\
$\ddagger$Purple Mountain Laboratories, Nanjing 211111, China\\
Email: 
\{zhiyu, chao\_feng, yong\_zeng, liteng, jinshi\}@seu.edu.cn} 
}
\maketitle

\begin{abstract}
  Wireless communication using fully passive metal reflectors is a promising technique for coverage expansion, signal enhancement, 
  rank improvement and blind-zone compensation, thanks to its appealing features including
  zero energy consumption, ultra low cost, signaling- and maintenance-free, easy deployment and full compatibility with existing and future wireless systems. 
  However, a prevalent understanding for reflection by metal plates is based on
  Snell's Law, i.e., signal can only be received when the observation angle equals to
  the incident angle, which is valid only when the electrical dimension of the metal plate is extremely large.
  In this paper, we rigorously derive a general reflection model that is applicable to metal reflectors of any size, any orientation, and any linear polarization. 
  The derived model is given compactly in terms of the radar cross section (RCS) of the metal plate, as a function of its physical dimensions and
  orientation vectors, as well as the wave polarization and the wave deflection vector, i.e., the change of direction from the incident wave direction to the observation direction. 
  Furthermore, experimental results based on actual field measurements are provided to validate the accuracy of our developed model and demonstrate the great potential of communications using metal reflectors.
\end{abstract}
\IEEEpeerreviewmaketitle
\section{Introduction}

\begin{figure}[!t]
  \setlength{\abovecaptionskip}{-0.1cm}
  \setlength{\belowcaptionskip}{-0.3cm}
  \centering
  \centerline{\includegraphics[width=2in,height=1.63398in]{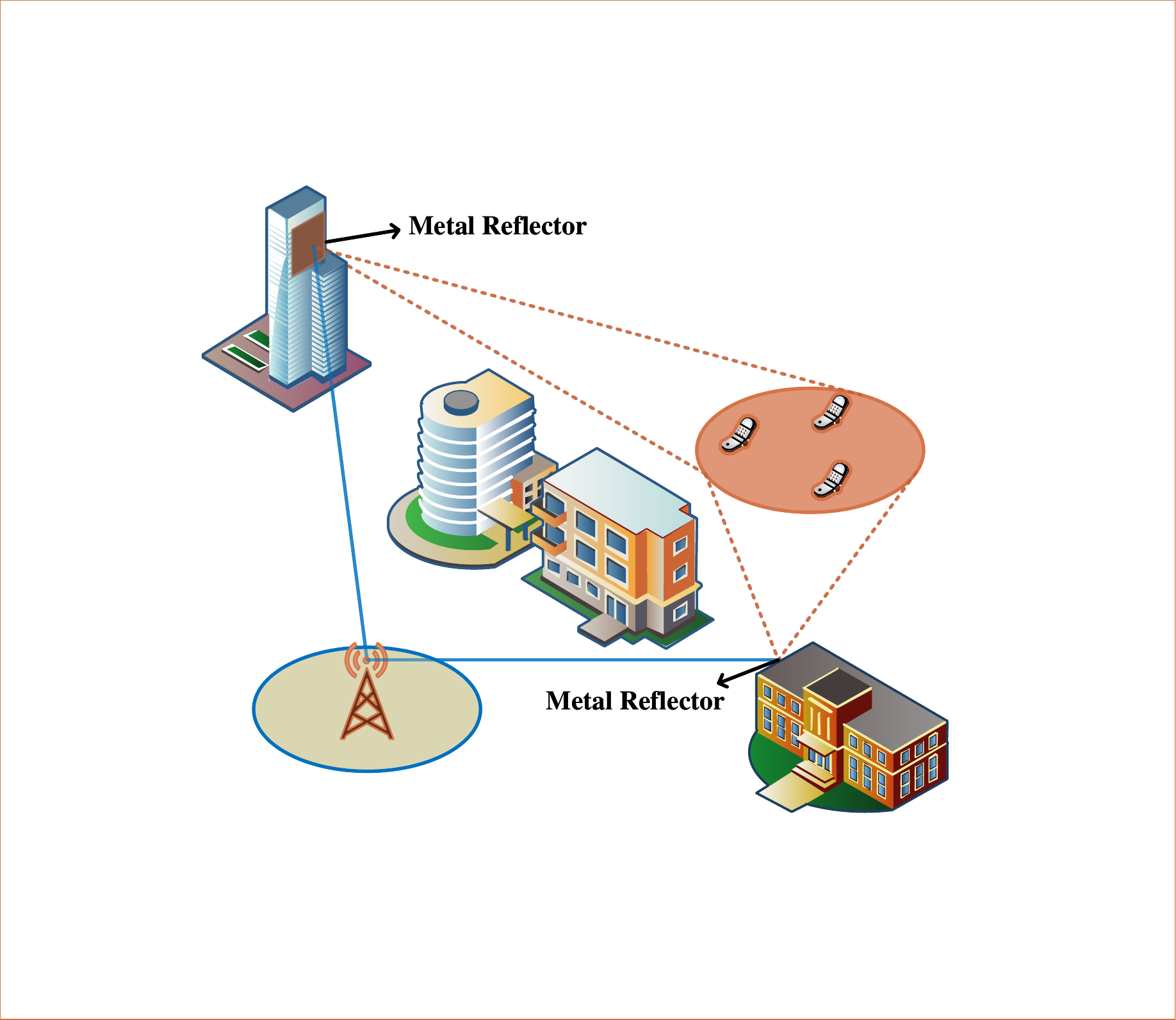}}
  \caption{Wireless communications using metal reflectors.}
  \label{scen}
  \vspace{-0.4cm}
\end{figure}

The history of wireless communication assisted by metal reflectors can be traced back to the 1960s, when metal reflectors were
used as passive relays in satellite communication systems due to their reliability and simplicity compared to active repeaters~\cite{cutler1965passive,ryerson1960passive,stahler1963corner}.
Furthermore, passive reflectors had also been exploited for antenna designs, with a wide range of applications including
radio astronomy, deep-space communication and satellite tracking~\cite{balanis2015antenna}\cite{RahmatSamii2014ReflectorA}. 
Recently, with the extensive exploration of beyond the fifth-generation (B5G) and the sixth-generation (6G) technologies, 
wireless communication using metal reflectors has gained renewed interest, since it provides a cost-effective solution for
coverage expansion, signal enhancement, rank improvement and blind-zone compensation, as illustrated in Fig.~\ref{scen}.
For instance, fully passive metal reflectors can be used for indoor communication improvement~\cite{huang2004investigation,barreiro2006passive,han2017enhancing}, 
millimeter-wave (mmWave) signal coverage enhancement~\cite{khawaja2018coverage,khawaja2019effect,khawaja2020coverage,el2022enhancement,peng2015effective,anjinappa2021base}, 
and energy redirection~\cite{romero2020irregular}.
Compared to the extensively studied semi-passive techniques like intelligent reflecting surfaces (IRSs) -assisted communications\cite{wu2019towards,basar2019wireless,ntontin2021optimal},
though metal reflectors cannot achieve dynamic reflection, 
they possess many attractive advantages, including zero energy consumption, ultra low cost,
signaling- and maintenance-free, easy deployment and full compatibility with existing and future wireless systems.

A prerequisite on the study of metal reflectors assisted communication is to properly model its reflected signal distribution in space. 
A common understanding for metal plate reflectors is that electromagnetic (EM) waves are merely reflected in the specular 
direction according to Snell's law of reflection~\cite{balanis2012advanced}, i.e., the reflecting angle should be equal to the incident angle, and no notable signal can be observed along other directions.
However, this turns out to be  
true only when the reflector size is much larger than the signal wavelength.
In practice, just like antennas, the reflected waves of a metal plate will actually constitute a beam, with beamwidth depending on its electrical dimensions.
This implies that though metal reflectors cannot dynamically adjust their reflected beam directions,
they can be effectively used to enhance the signal coverage of dedicated areas, as illustrated in Fig.~\ref{scen}.
There have been some preliminary efforts on the mathematical modelling of the signal reflection by passive or semi-passive reflectors~\cite{balanis2012advanced,ozdogan2019intelligent,najafi2020physics}.
For instance, by considering IRS-assisted communications, 
the reflection models of IRS were derived in~\cite{ozdogan2019intelligent} and\cite{najafi2020physics}, which are also applicable to metal reflectors.
However, such existing models were derived based on angles defined with respect to the fixed normal vector  of the plate, which makes it very difficult to analyze
the impact of plate rotation or to be utilized for the deployment optimization of metal plates. Furthermore, no experimental measurements were provided to verify the existing models.  

To address the above issues, in this paper, we develop a rigorous and generic reflection model for metal reflectors,
which is applicable to metal plates of any size, any orientation, and any linear polarization. 
The derived model is given compactly in terms of the radar cross section (RCS) of the metal plate, as a function of its
physical dimensions and orientation vectors, as well as the wave polarization and the deflection vector, i.e., 
the change of direction from the incident wave direction to the observation direction.
To gain more insights, we further study some special cases of our developed model, 
considering typical polarizations that include the existing results as special cases~\cite{balanis2012advanced,ozdogan2019intelligent,najafi2020physics}. 
Furthermore, experimental results based on field measurements are provided to validate the accuracy of our developed model and demonstrate the great potential of wireless communications using metal reflectors.

\section{System model}
 \begin{figure}[!t]
  \setlength{\abovecaptionskip}{-0.1cm}
  \setlength{\belowcaptionskip}{-0.3cm}
  \centering
  \centerline{\includegraphics[scale=0.15]{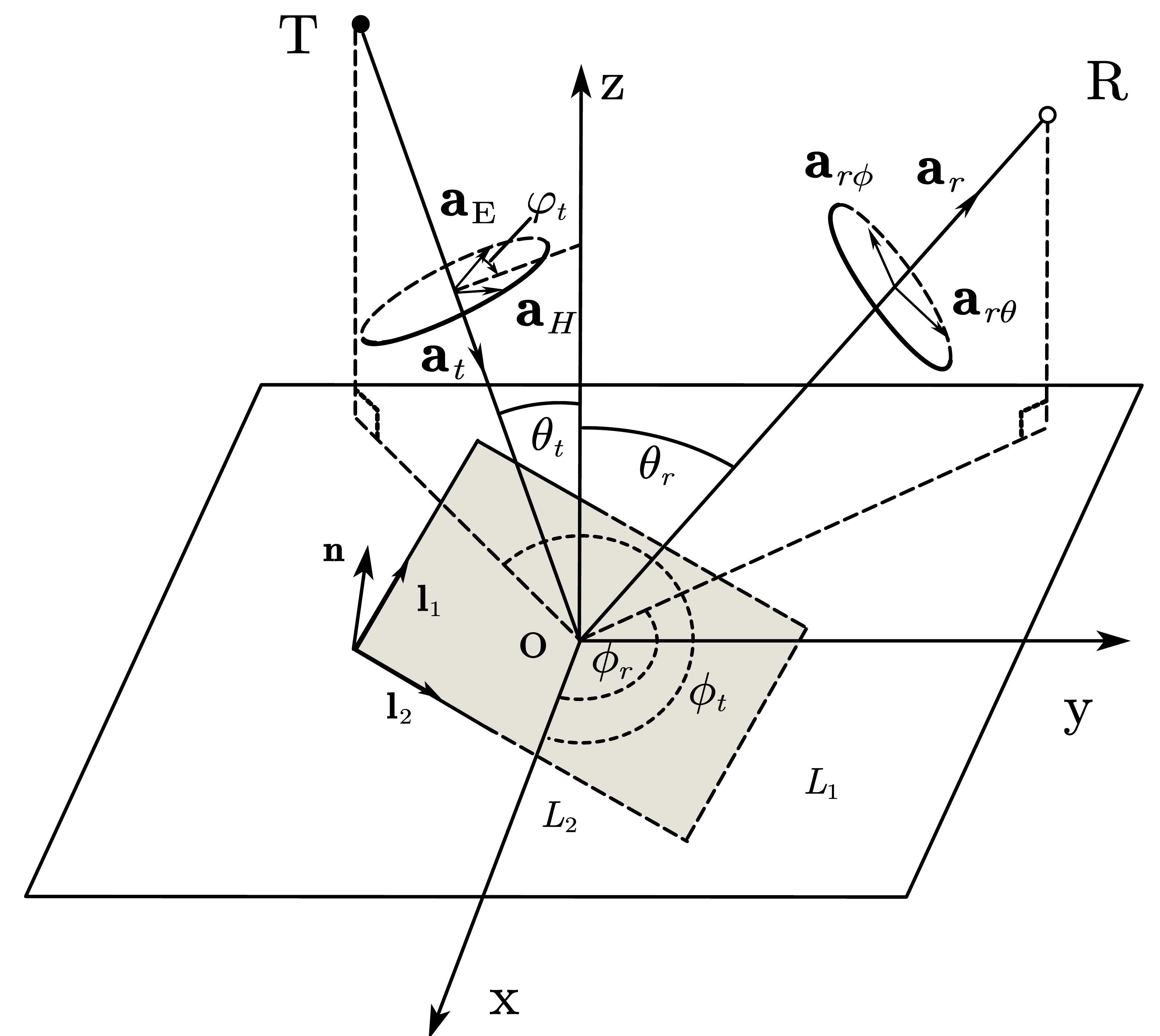}}
  \caption{Wireless communication assisted by a metal plate reflector with arbitrary orientation.}
  \label{reflectionedmodel}
  \vspace{-0.3cm}
\end{figure}

As shown in Fig.~\ref{reflectionedmodel}, we consider a wireless communication system assisted by a rectangular metal plate reflector that may have an arbitrary orientation.
Without loss of generality, a three-dimensional Cartesian coordinate system 
is chosen so that the plate center locates at the origin.
Note that different from the existing works~\cite{ozdogan2019intelligent}\cite{najafi2020physics}, where the plane of the metal plate is chosen as the $x$-$y$ plane, 
we consider a more general scenario that the metal plate may be tilted relative to the $x$-$y$ plane, and its normal vector is denoted as $\mathbf{n}$.
This makes our derived results readily used to analyze the impact of plate orientation and optimize the plate deployment.
The physical size of the metal plate is denoted as $L_1 \times L_2$,
with $\mathbf{l}_1$ and $\mathbf{l}_2$ being the unit vectors denoting the directions of its two edges. Obviously,
$\mathbf{n}$, $\mathbf{l}_1$, and $\mathbf{l}_2$ are mutually orthonormal vectors, and they determine the orientation of the metal plate.

A uniform plane wave (UPW) originated from point $T$ impinges on the metal plate, with its incident direction denoted by vector ${\bf{a}}_{t}$.
The incident wave can be expressed by its electric and magnetic field distributions, i.e.,
\begin{equation}\label{incdientwavee}
  {{\bf{E}}^t = \eta H_{0} e^{-j k {\bf{a}}_{t} \cdot\left({\bf{e}}_{x} x+{\bf{e}}_{y} y+{\bf{e}}_{z} z\right)} {\bf{a}}_{E}},
\end{equation}
\begin{equation}\label{incdientwaveh}
  {{\bf{H}}^t = H_{0} e^{-j k {\bf{a}}_{t} \cdot\left({\bf{e}}_{x} x+{\bf{e}}_{y} y+{\bf{e}}_{z} z\right)} {\bf{a}}_{H}},
\end{equation}
where $H_0$ is the magnitude of the incident magnetic field and $\eta$ is the characteristic impedance;
$j$ is the imaginary unit with $j^2=-1$; 
$k=\frac{2\pi}{\lambda}$ denotes the wave number with wavelength $\lambda$;
${\bf{e}}_{x}$, ${\bf{e}}_{y}$, ${\bf{e}}_{z}$ denote the unit vectors along $x$, $y$ and $z$ directions, respectively;
${\bf{a}}_{E}$ and ${\bf{a}}_{H}$ denote the direction vectors of the electric and magnetic fields, respectively.
For UPW, we have ${\bf{a}}_t={\bf{a}}_E\times{\bf{a}}_H$, with ``$\times$" denoting the cross product of two vectors. Furthermore, the dot notation ``$\cdot$"
in~\eqref{incdientwavee} and~\eqref{incdientwaveh} denotes the inner product of two vectors.

By treating the metal plate as a perfect electric conductor and applying physical optics techniques, 
the current density ${\bf{J}}_s$ induced at the point ${\bf{r}}^{\prime }$ on the metal plate 
can be expressed as~\cite{balanis2012advanced}
\begin{equation}\label{js}
  {\bf{J}}_s=2 {\bf{n}} \times {\bf{H}}^t=2H_0 {\bf{n}}\times{{\bf{a}}_H} e^{-jk{\bf{a}}_t\cdot{\bf{r}}^{\prime }}.
\end{equation}

Obviously, the induced current flows in the plane occupied by the metal plate.
We aim to rigorously derive the reflected signal strength at any observation point $R$, whose direction vector is 
denoted as $\mathbf{a}_r$, as shown in Fig.~\ref{reflectionedmodel}.
When $R$ is located in the far-field region of the metal plate, the reflected electric field can be expressed by its spherical components, given by~\cite{balanis2012advanced}
\begin{equation}
  E^{r}_{\rho} \simeq 0,
\end{equation}
\begin{equation}\label{etheta}
  E^{r}_{\theta} \simeq-\frac{j k \eta e^{-j k d_r}}{4 \pi d_r} \iint_{S} {\bf{J}}_s \cdot {\bf{a}}_{r\theta} e^{+j k {\bf{r}}^{\prime } \cdot {\bf{a}}_r} {\rm d} s,
\end{equation}
\begin{equation}\label{ephi}
  E^{r}_{\phi} \simeq-\frac{j k \eta e^{-j k d_r}}{4 \pi d_r} \iint_{S} {\bf{J}}_s \cdot {\bf{a}}_{r\phi} e^{+j k {\bf{r}}^{\prime } \cdot {\bf{a}}_r} {\rm d} s,
\end{equation}
where $S$ denotes surface occupied by the metal plate; $d_r$ is the distance between the origin 
and the observation point $R$; 
${\bf{a}}_{r\theta}$ and ${\bf{a}}_{r\phi}$ 
are orthonormal vectors denoting the Cartesian-to-spherical component transformation.

\section{The Derived reflection Model}
In this section, we derive the reflection model of the rectangular metal plate based on \eqref{etheta} and \eqref{ephi}, in terms of its RCS as a function of 
plate's physical dimensions ($L_1$, $L_2$), orientation vectors (${\bf{n}}$, ${\bf{l}}_1$, ${\bf{l}}_2$), magnetic field direction vector ${\bf{a}}_H$, as well as the wave deflection vector ${\bf{a}}_r-{\bf{a}}_t$. 
Furthermore, some special cases of the derived model are studied to gain useful insights.

\subsection{RCS of Plate Reflectors}
According to~\cite{balanis2015antenna}, RCS is defined as the area intercepting the amount of power that, when scattered isotropically, 
produces at the receiver a density that is equal to the density scattered by the actual target.
Once the RCS $\sigma$ is known, the ratio of the received signal power to the transmitted power after scattering/reflecting by the object can be expressed as~\cite{balanis2015antenna}
\begin{equation}\label{pr}
  \frac{P_{r}}{P_{t}}=\frac{G_{t} G_{r} \sigma \lambda^{2}}{4 \pi\left(4 \pi d_{t} d_{r}\right)^{2}},
\end{equation}
where $G_t$ is the gain of the transmitting antenna; $G_r$ is the gain of the receiving antenna;
$d_t$ is the distance from the transmitter to the reflector, and $d_r$ is the distance from the reflector to the receiver. 
Therefore, when all other parameters are fixed,
RCS provides a direct characterization for the reflection model of metal plates. 

\begin{theorem}
  \label{RCS}
  The RCS of the rectangular metal plate in Fig.~\ref{reflectionedmodel} when observed at direction ${\bf{a}}_r$, for an incident UPW with direction ${\bf{a}}_t$ and magnetic field direction ${\bf{a}}_H$, can be expressed as 
  \begin{align}\label{sigma}
    &\qquad \qquad \sigma = \underbrace{\frac{4\pi L_1^2 L_2^2 }{{\lambda}^2}}_{\sigma_{\max}}   \underbrace{  ||  \left( {\bf{n}} \times {\bf{a}}_H \right) \times {\bf{a}}_r ||^2 }_{f_{J_s}\leq 1} \\ 
     &\times \underbrace{ \sinc^2 \left( \frac{k L_1}{2} \left({\bf{a}}_r-{\bf{a}}_t \right) \cdot {\bf{l}}_1 \right)   \sinc^2 \left( \frac{k L_2}{2} \left({\bf{a}}_r-{\bf{a}}_t \right) \cdot {\bf{l}}_2 \right).}_{f_{AF}\leq 1}\nonumber
  \end{align}
\end{theorem}

\begin{IEEEproof}
  Please refer to Appendix A.
\end{IEEEproof}

\begin{figure}[!t]
  \setlength{\abovecaptionskip}{-0.1cm}
  \setlength{\belowcaptionskip}{-0.3cm}
  \centering
  \centerline{\includegraphics[scale=0.15]{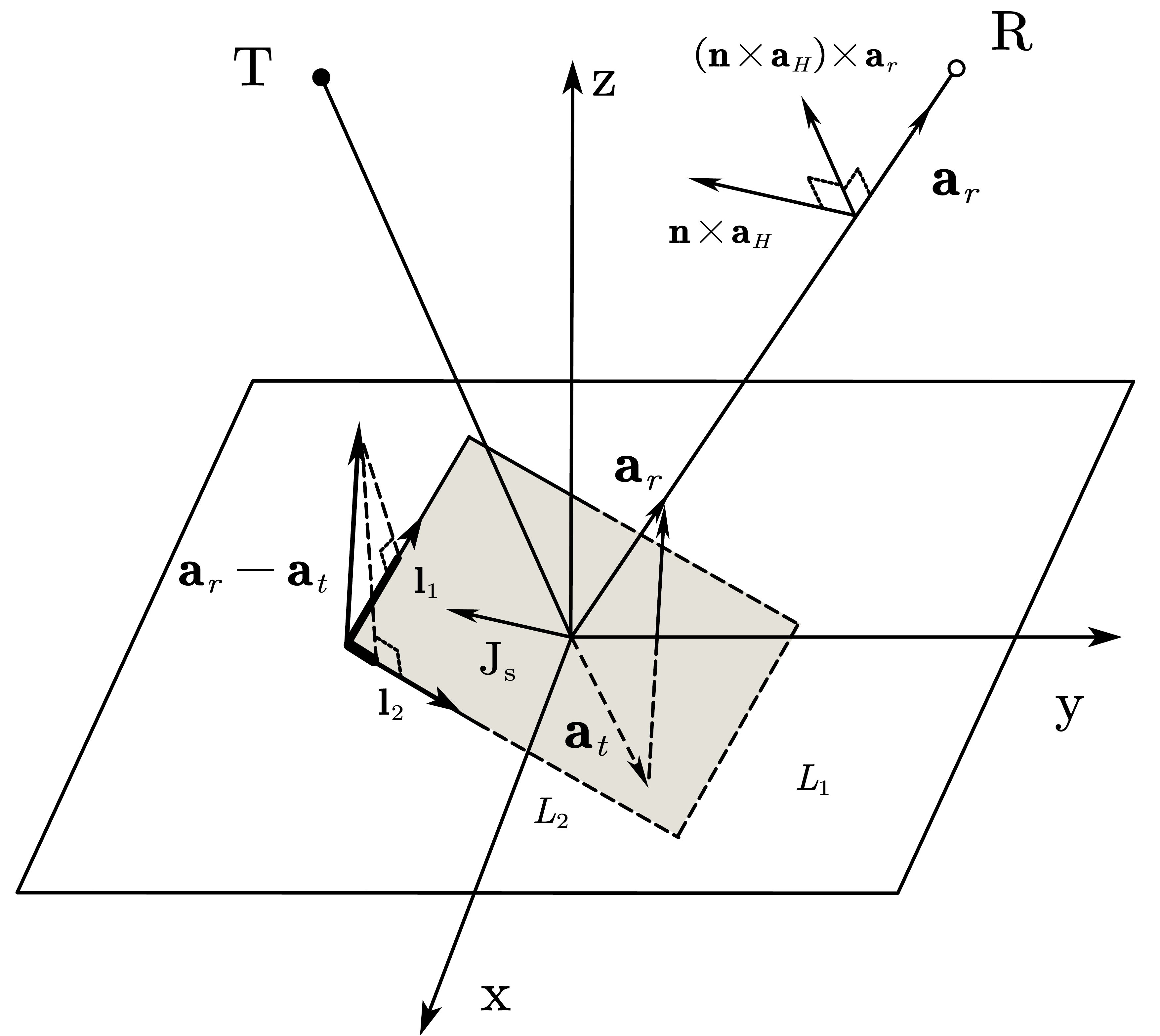}}
  \caption{Illustrations for $f_{J_s}$ and $f_{AF}$.}
  \label{modelofinsight}
  \vspace{-0.3cm}
\end{figure}

Some important observations can be made from Theorem~\ref{RCS}. 
Firstly, the RCS can be decomposed into three components, namely the maximum possible RCS value $\sigma_{\max}$, 
and two factors $f_{Js}$ and $f_{AF}$ of values no greater than 1.  
Secondly, the maximum RCS $\sigma_{\max}$ is proportional to the square of the plate area $A=L_1L_2$. 
This is expected since a larger plate not only intercepts more energy for signal reception, 
but also forms a narrower beam for signal reflection.
Thirdly, the factor $f_{Js}$ depends on the square of the length of $\left( {\bf{n}} \times {\bf{a}}_H \right) \times {\bf{a}}_r$.
Lastly, analogously to antenna arrays, 
the factor $f_{AF}$ can be interpreted as the array factor, whose beamwidth depends on the electrical dimensions $\frac{L_1}{\lambda}$ and $\frac{L_2}{\lambda}$.
Furthermore,  
$\left({\bf{a}}_r-{\bf{a}}_t \right) \cdot {\bf{l}}_1$ and $\left({\bf{a}}_r-{\bf{a}}_t \right) \cdot {\bf{l}}_2$ 
correspond to the projection of the deflection vector $\left({\bf{a}}_r-{\bf{a}}_t \right)$ along the two edges of the plate ${\bf{l}}_1$ and ${\bf{l}}_2$, respectively, as illustrated in Fig.~\ref{modelofinsight}.

\begin{lemma}\label{lemma1}
The array factor $f_{AF}$ achieves its maximum value of $f_{AF}=1$ 
at the direction of specular reflection, i.e.,
when $({\bf{a}}_r-{\bf{a_t}})\parallel {\bf{n}}$, or equivalently  ${\bf{a}}_r={\bf{a}}_t-2\left({\bf{n}}\cdot{\bf{a}}_t\right){\bf{n}}$.
Furthermore, when $L_1,L_2\gg \lambda$, the RCS in \eqref{sigma} reduces to
\begin{equation}
  \begin{aligned}
    &\sigma=\\
    &
    \begin{cases}
      \!\sigma_{\max}|| \! \left( \!{\bf{n}}\! \times \! {\bf{a}}_H \! \right) \! \times \!\left( \!{\bf{a}}_t\!-\!2\!\left({\bf{n}}\!\cdot\!{\bf{a}}_t\right){\bf{n}} \!\right)\! ||^2&,\text{if }{\bf{a}}_r\!=\!{\bf{a}}_t\!-\!2\!\left(\!{\bf{n}}\!\cdot\!{\bf{a}}_t\!\right)\!{\bf{n}}\\
      0&,\text{otherwise}
    \end{cases}.
  \end{aligned}
\end{equation}
\end{lemma}
\begin{IEEEproof}
  It is obviously observed that the maximum $f_{AF}$ occurs only when $({\bf{a}}_r-{\bf{a_t}}) \cdot {\bf{l}}_1=0$ and $({\bf{a}}_r-{\bf{a_t}}) \cdot {\bf{l}}_2=0$,
  and we can easily obtain $({\bf{a}}_r-{\bf{a_t}})\parallel {\bf{n}}$, or equivalently ${\bf{a}}_r={\bf{a}}_t-2\left({\bf{n}}\cdot{\bf{a}}_t\right){\bf{n}}$.
  Besides, when $L_1,L_2\gg \lambda$, $f_{AF}$ reduces to $0$ if ${\bf{a}}_r\neq {\bf{a}}_t-2\left({\bf{n}}\cdot{\bf{a}}_t\right){\bf{n}}$.
\end{IEEEproof}

Lemma~\ref{lemma1} shows that our derived expression~\eqref{sigma} includes the well-known Snell's law of reflection as a special case, i.e., 
when the plate dimensions are much larger than signal wavelength, signal can only be received in the specular direction. 
\subsection{Some special cases}

Different from existing models like~\cite{balanis2012advanced,ozdogan2019intelligent,najafi2020physics}, 
our newly derived model in Theorem~\ref{RCS} is given compactly in vector forms. This makes it very generic and convenient for use.  
To gain more insights,
we further present the results from~\eqref{sigma} in terms of incident, observation, and polarization angles.
As shown in Fig.~\ref{reflectionedmodel}, let $\theta_t\in \left[\right.0, \frac{\pi}{2} \left.\right)$ and $\phi_t\in \left[\right.0, 2\pi\left.\right)$
denote the zenith and azimuth angles of the incident wave, respectively.
$\varphi_{t}\in\left(\right.0,2\pi\left.\right]$ denotes the polarization angle, 
which is defined as the angle between the electric field direction ${\bf{a}}_E$ and the reference plane formed by $z$-axis and the incident direction ${\bf{a}}_t$;
$\theta_r\in \left[\right.0, \frac{\pi}{2} \left.\right)$ and $\phi_r\in \left[\right.0, 2\pi\left.\right)$
denote the zenith and azimuth angles of the observation direction, respectively.
Therefore, the direction vectors ${\bf{a}}_{t}$, ${\bf{a}}_{r}$ and ${\bf{a}}_{H}$ can be expressed in terms of the above angles as (See Appendix B for the proof)
\vspace{-0.2cm}
\begin{equation}\label{defineat}
  {\bf{a}}_{t}=\left(-\sin \theta_{t} \cos \phi_{t},-\sin \theta_{t} \sin \phi_{t},-\cos \theta_{t}\right),
\end{equation}
\begin{equation}\label{definear}
  {\bf{a}}_{r}=\left(\sin \theta_{r} \cos \phi_{r},\sin \theta_{r} \sin \phi_{r},\cos \theta_{r}\right),
\end{equation}
\begin{equation}\label{defineah}
  \begin{aligned}
    &\quad{\bf{a}}_{H}=\left(-\sin \varphi_{t} \cos \theta_{t} \cos \phi_{t}-\cos \varphi_{t} \sin \phi_{t},\right. \\
    \quad&\left.-\sin \varphi_{t} \cos \theta_{t} \sin \phi_{t}+\cos \varphi_{t} \cos \phi_{t}, \sin \varphi_{t} \sin \theta_{t}\right).\\
  \end{aligned}
\end{equation}

\begin{figure}[!t]
  \setlength{\abovecaptionskip}{-0.1cm}
  \setlength{\belowcaptionskip}{-0.3cm}
  \centering
  \centerline{\includegraphics[scale=0.15]{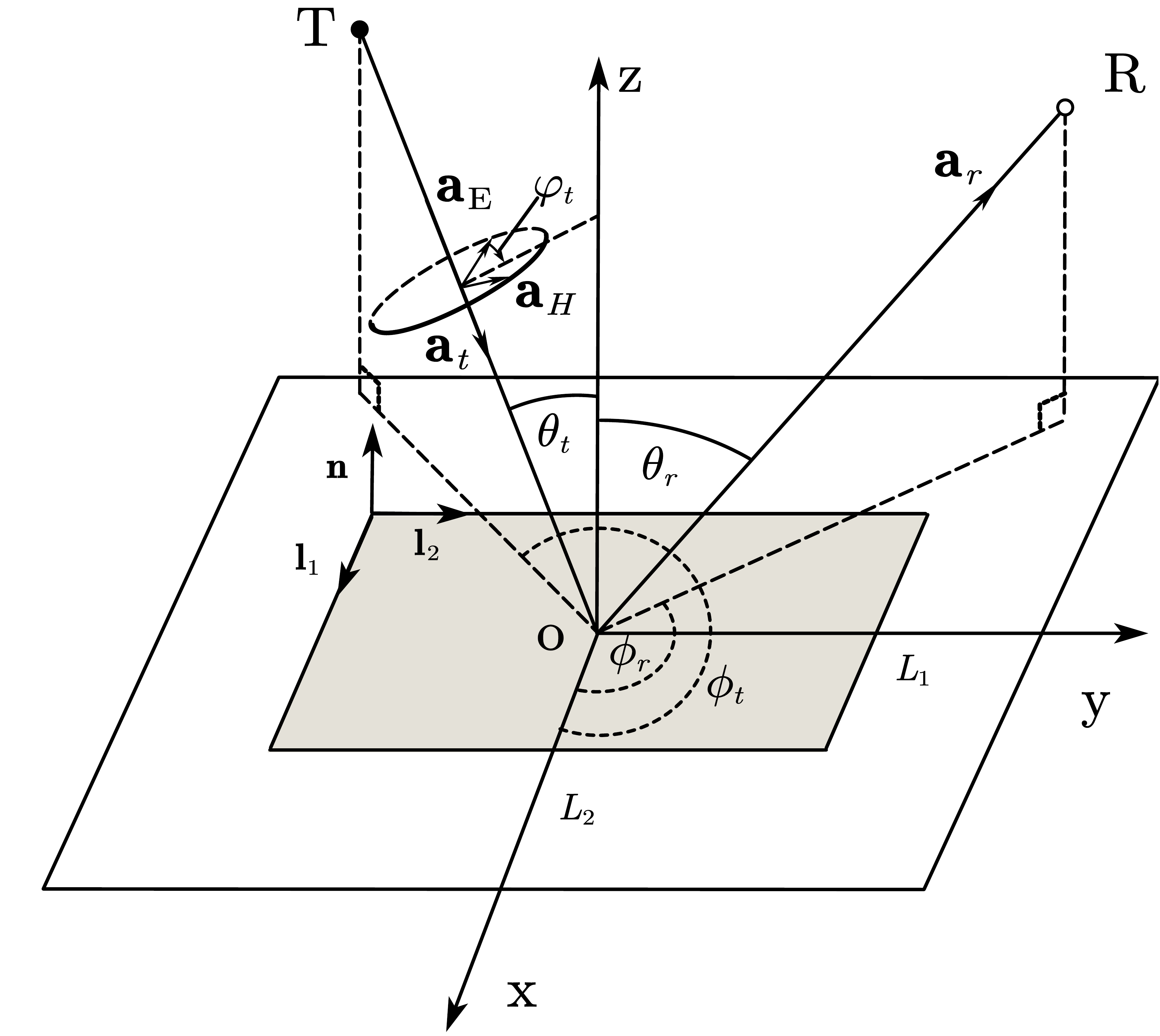}}
  \caption{Special case where the metal plate is placed on $x$-$y$ plane with the two edges parallel to $x$- and $y$-axis, respectively.}
  \label{simplifiedcases}
  \vspace{-0.4cm}
  \end{figure}

\begin{lemma}
  \label{SIMPLERCS}
  For the special case when the metal plate is placed 
  on the $x$-$y$ plane 
  with its two edges parallel to $x$- and $y$-axis, as shown in Fig.~\ref{simplifiedcases}, the general result in~\eqref{sigma} reduces to~\eqref{eqs1} shown at the top of the next page.
\end{lemma}

\begin{IEEEproof}
  The proof can be readily obtained by noting 
  ${\bf{n}}={\bf{e}}_z,{\bf{l}}_1 ={\bf{e}}_x,{\bf{l}}_2 ={\bf{e}}_y$, 
  and substituting~\eqref{defineat}-\eqref{defineah} into~\eqref{sigma}.
\end{IEEEproof}

Note that 
\eqref{eqs1} is consistent with the existing result in~\cite{najafi2020physics}, except for some differences in the definition of symbols.
To further simplify the results in \eqref{eqs1}, we consider two special polarizations in the following.

  \begin{figure*}[ht] 
    \centering
    \begin{equation}\label{eqs1}	    
       \begin{aligned}
        \sigma&=\sigma_{\max}
        \left[
          \cos^2 \theta_{r}\left (\sin \varphi_{t} \cos \theta_{t} \sin \left(\phi_{r}-\phi_{t}\right)+\cos \varphi_{t}  \cos\left(\phi_{t}-\phi_{r}\right)\right)^{2}
          +\left(\cos \varphi_{t} \sin \left(\phi_{t}-\phi_{r}\right)+\sin \varphi_{t} \cos \theta_{t} \cos \left(\phi_{t}-\phi_{r}\right)\right)^{2}
        \right]\\
       & \times\operatorname{\sinc}^2 \left(\frac{k L_{1}}{2}\left(\sin \theta_{r} \cos \phi_{r}+\sin \theta_{t} \cos \phi_{t}\right)\right)  \operatorname{\sinc}^2 \left(\frac{k L_{2}}{2}\left(\sin \theta_{r} \sin \phi_{r}+\sin \theta_{t} \sin \phi_{t}\right)\right)  \\
       \end{aligned}
    \end{equation}
    \hrulefill
    \vspace*{4pt}
  \end{figure*}
\begin{corollary}
  \label{lemma special case 1.1}
  For 
  $\varphi_t=\frac{\pi}{2}$ or $\frac{3\pi}{2}$ and $\phi_t=\frac{3\pi}{2}$,
  the RCS in \eqref{eqs1} reduces to 
  \begin{equation}\label{eqs1.1}
    \begin{split}
      \sigma&=\sigma_{\max}\left[\left(\cos \theta_{t} \cos \theta_{r} \cos \phi_{r}\right)^{2}+\left(\cos \theta_{t} \sin \phi_{r}\right)^{2}\right]\\
      & \times \operatorname{\sinc}^2 \left(\frac{k L_{1}}{2}\sin \theta_{r} \cos \phi_{r}\right)\\
      &  \times \operatorname{\sinc}^2 \left(\frac{k L_{2}}{2}\left(\sin \theta_{r} \sin \phi_{r}-\sin \theta_{t} \right)\right).
    \end{split}
  \end{equation}

  Note that the specialized expression~\eqref{eqs1.1} is consistent with the existing result in \cite{balanis2012advanced}.
  Furthermore, if $\varphi_r=\frac{\pi}{2}$,
  we have
  \begin{equation}
    \label{case 1.1.1}
    \sigma=\sigma_{\max} \cos ^{2} \theta_{t} \operatorname{\sinc}^{2}\left(\frac{k L_{2}}{2} \left(\sin \theta_{r}-\sin \theta_{t}\right)\right).
  \end{equation}
\end{corollary}

\begin{corollary}
  \label{lemma special case 1.2}
  For $\varphi_t=0$ or $\pi$ and $\phi_t=\frac{3\pi}{2}$,
  the RCS in~\eqref{eqs1} reduces to 
  \begin{equation}\label{eqs1.2}
    \begin{split}
      \sigma&=\sigma_{\max}\left[\left(\cos \theta_{r} \sin \phi_{r}\right)^{2}+\left(\cos \phi_{r}\right)^{2}\right]\\
      & \times\operatorname{\sinc}^2 \left(\frac{k L_{1}}{2}\sin \theta_{r} \cos \phi_{r}\right)\\
      & \times  \operatorname{\sinc}^2 \left(\frac{k L_{2}}{2}\left(\sin \theta_{r} \sin \phi_{r}-\sin \theta_{t} \right)\right).
    \end{split}
  \end{equation}
 
  Furthermore, if $\varphi_r=\frac{\pi}{2}$,
  we have 
  \begin{equation}\label{case 1.2.1}
    \sigma=\sigma_{\max} \cos ^{2} \theta_{r}  \operatorname{\sinc}^{2}\left(\frac{k L_{2}}{2}\left(\sin \theta_{r}-\sin \theta_{t}\right)\right).
  \end{equation}
\end{corollary}

It is observed from \eqref{case 1.1.1} and \eqref{case 1.2.1},
that the maximum reflection direction of metal plates 
also depends on the polarization angle of the incident wave,
rather than always occurs exactly 
at the specular reflection direction with $\theta_r=\theta_t$.

\section{Experimental Verification}
\begin{figure}[!t]
  \setlength{\abovecaptionskip}{-0.1cm}
  \setlength{\belowcaptionskip}{-0.1cm}
  \centering
  \centerline{\includegraphics[scale=0.30]{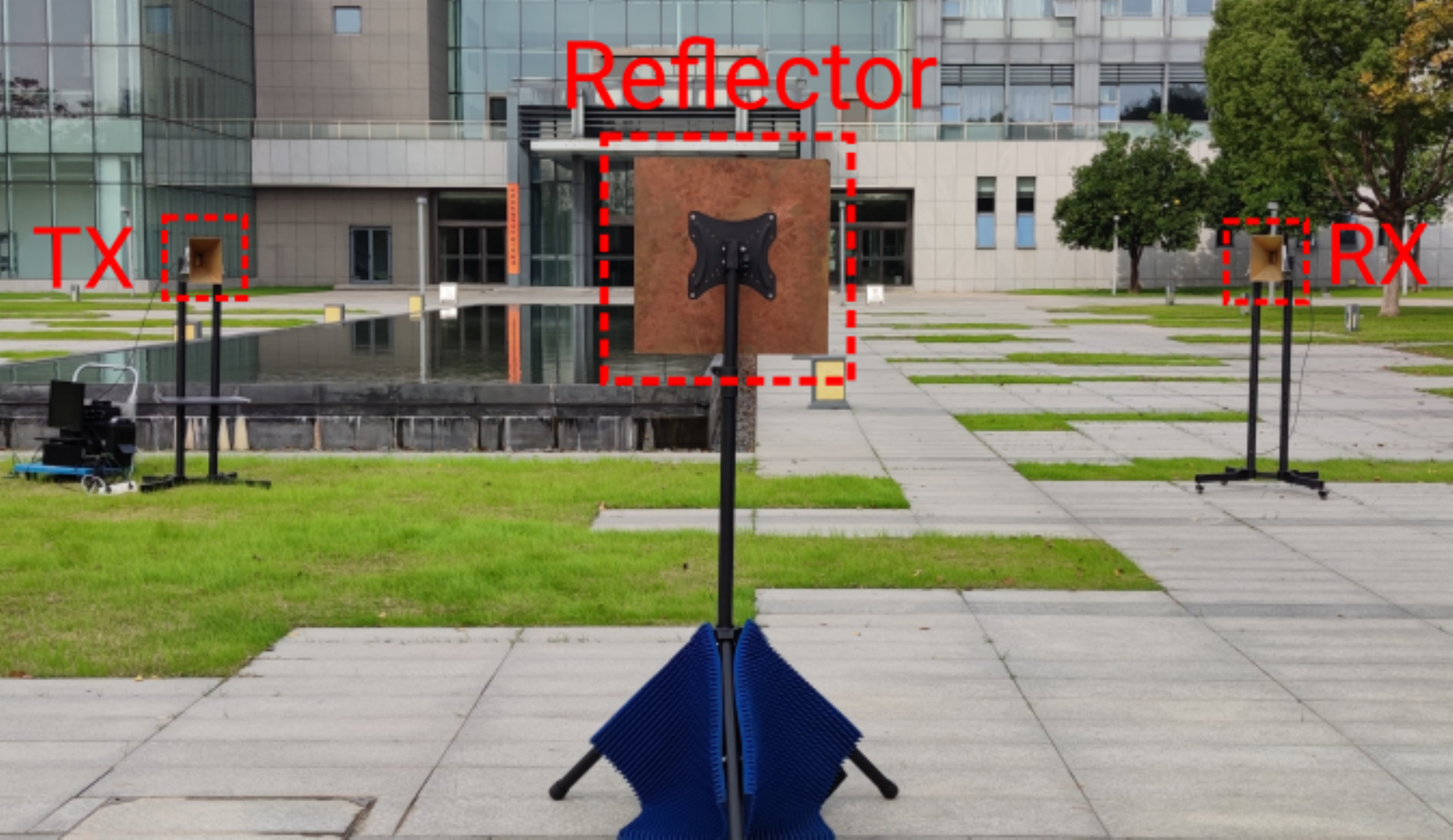}}
  \caption{Measurement scenario viewed from the metal reflector.}
  \label{measurement scenario1}
\end{figure}
\begin{figure}[!t]
  \setlength{\abovecaptionskip}{-0.1cm}
  \setlength{\belowcaptionskip}{-0.3cm}
  \centering
  \centerline{\includegraphics[scale=0.20]{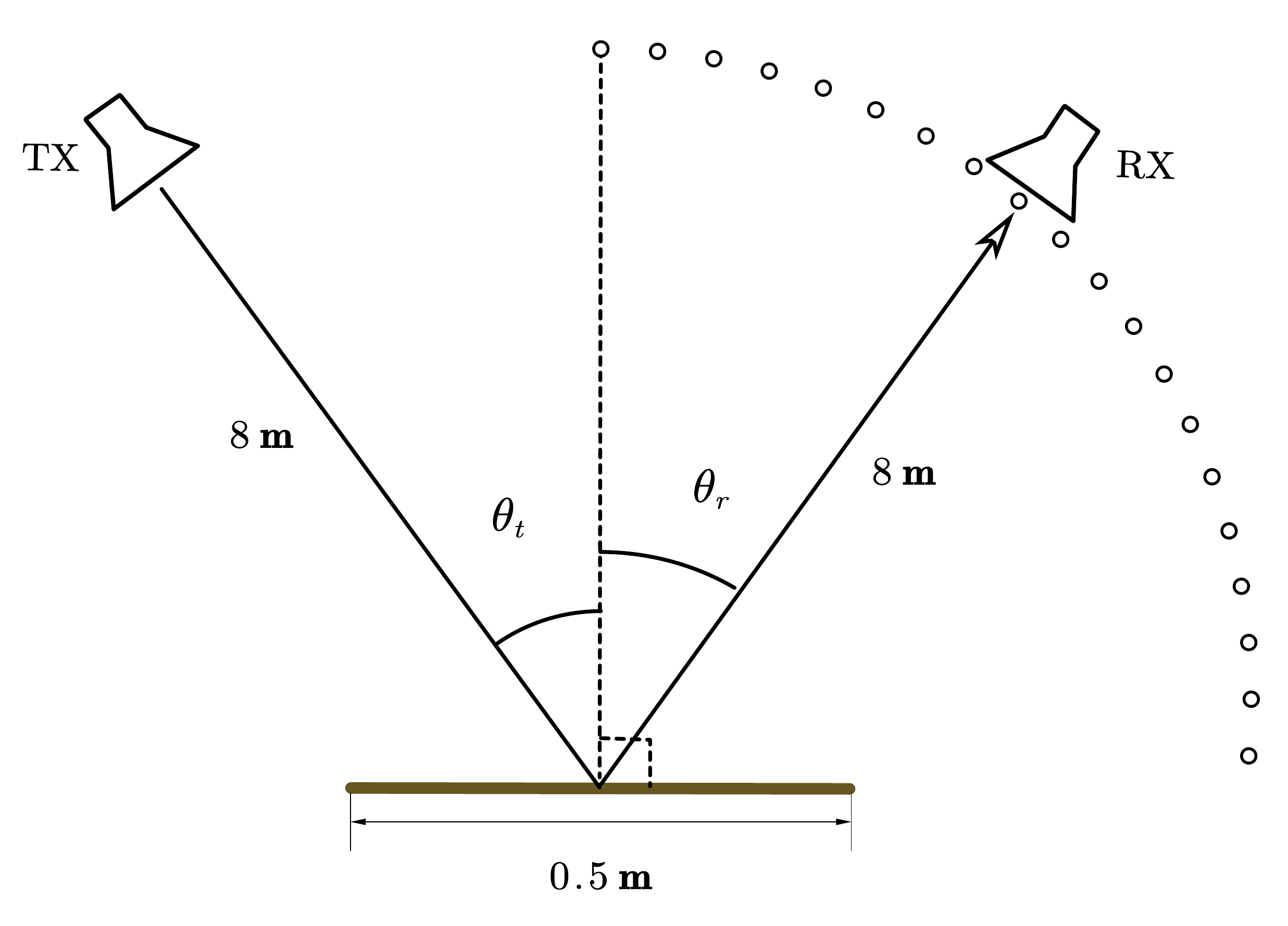}}
  \caption{The geometric setup for the experimental measurement.}
  \label{measurement scenario2}
\end{figure}
In this section, we validate our developed model with experimental measurements in an open space. 
For convenience of measurement, the models~\eqref{case 1.1.1} and~\eqref{case 1.2.1} are verified. 
We use Universal Software Radio Peripheral (USRP) 2974, 
together with a power amplifier and a horn antenna (TX) to transmit a continuous single-tone EM wave at $3\,{\mathrm{GHz}}$.
A handheld RF analyzer (KEYSIGHT N9914A), 
together with a horn antenna (RX),
is used to measure the received power for different incident and observation angles, as shown in Fig.~\ref{measurement scenario1} and Fig.~\ref{measurement scenario2}. 
The transmitting and receiving antennas and the center of the metal reflector are placed at a height of $1.5\,{\mathrm{m}}$ above the ground.
The TX is fixed and oriented towards the center of 
the reflector with incident angle $\theta_t=25\,{\mathrm{^{\circ}}}$, $45\,{\mathrm{^{\circ}}}$ and $65\,{\mathrm{^{\circ}}}$.
The RX is placed at 19 positions along 
a quarter of a circular arc from $\theta_r=0\,{\mathrm{^{\circ}}}$ to $\theta_r=90\,{\mathrm{^{\circ}}}$ with an interval of $5\,{\mathrm{^{\circ}}}$.
Throughout the measurement, the orientations of the transmitting
and receiving antennas are both aligned to the center of the metal plate. 
The metal plate is made of copper with size $L_1=L_2=5\lambda$, and the thickness of the plate is $0.5\,{\mathrm{cm}}$.
The distance from the TX to the center of the metal plate and that from the center of the metal plate to the RX are both $8\,{\mathrm{m}}$. 
The transmitting power of the USRP is $0\,{\mathrm{dBm}}$, and the gain of the power amplifier is $38.861\,{\mathrm{dB}}$.
The Half-Power Beam Width (HPBW) of the two horn antennas in $E$ and $H$ planes are $33.31\,{\mathrm{^{\circ}}}$ and $30.81\,{\mathrm{^{\circ}}}$, respectively.
The gains of the transmitting and receiving antennas are about $16\,{\mathrm{dBi}}$. 
The measurement parameters are summarized in Table~\ref{MEASUREMENT PARAMETERS}. 

\begin{figure*}[htbp]
  \centering
  \subfigure[$\theta_t=25^{\circ} $]{\includegraphics[width=1.98502in,height=1.5in]{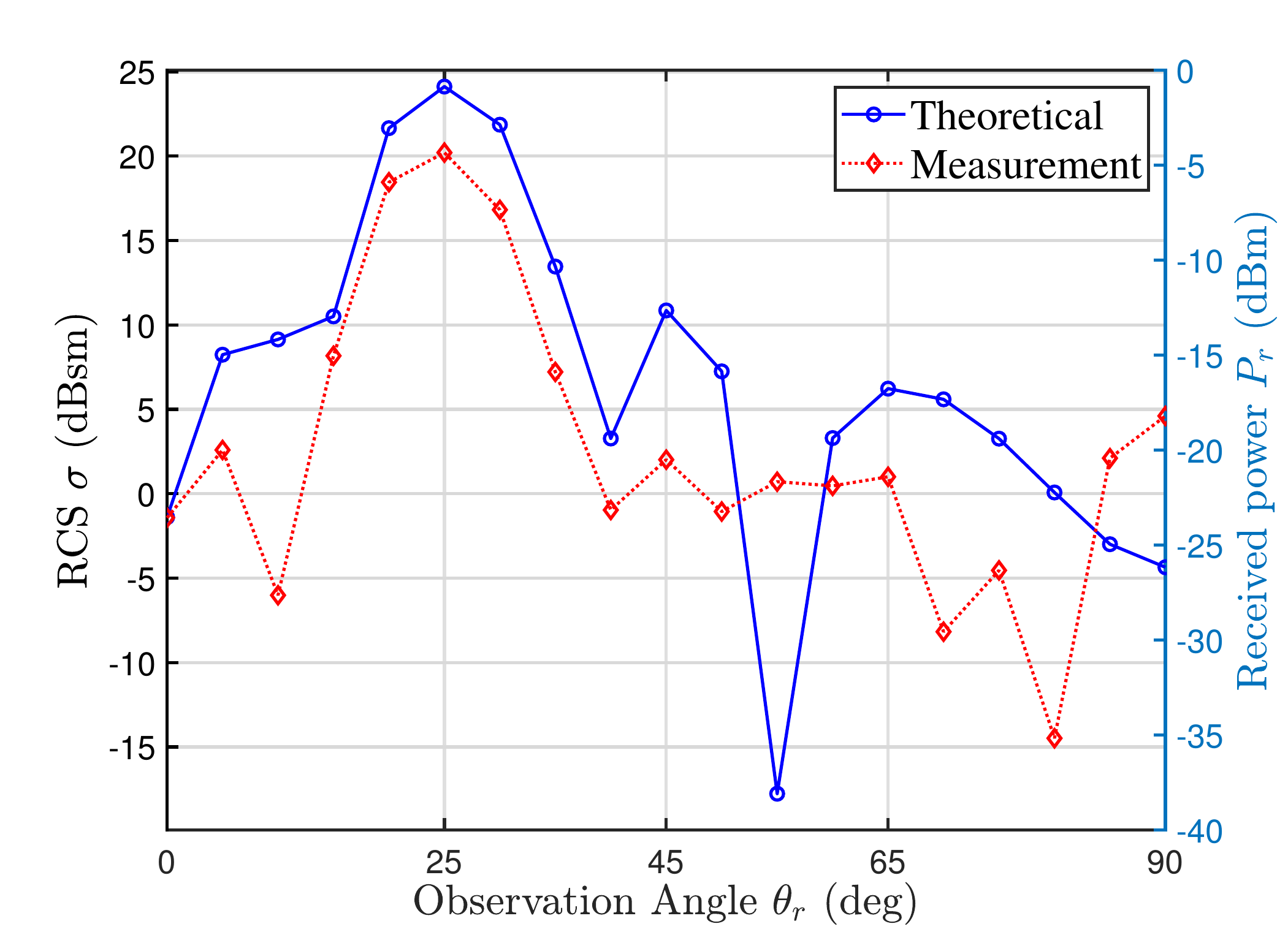}}
  \subfigure[$\theta_t=45^{\circ} $]{\includegraphics[width=1.98502in,height=1.5in]{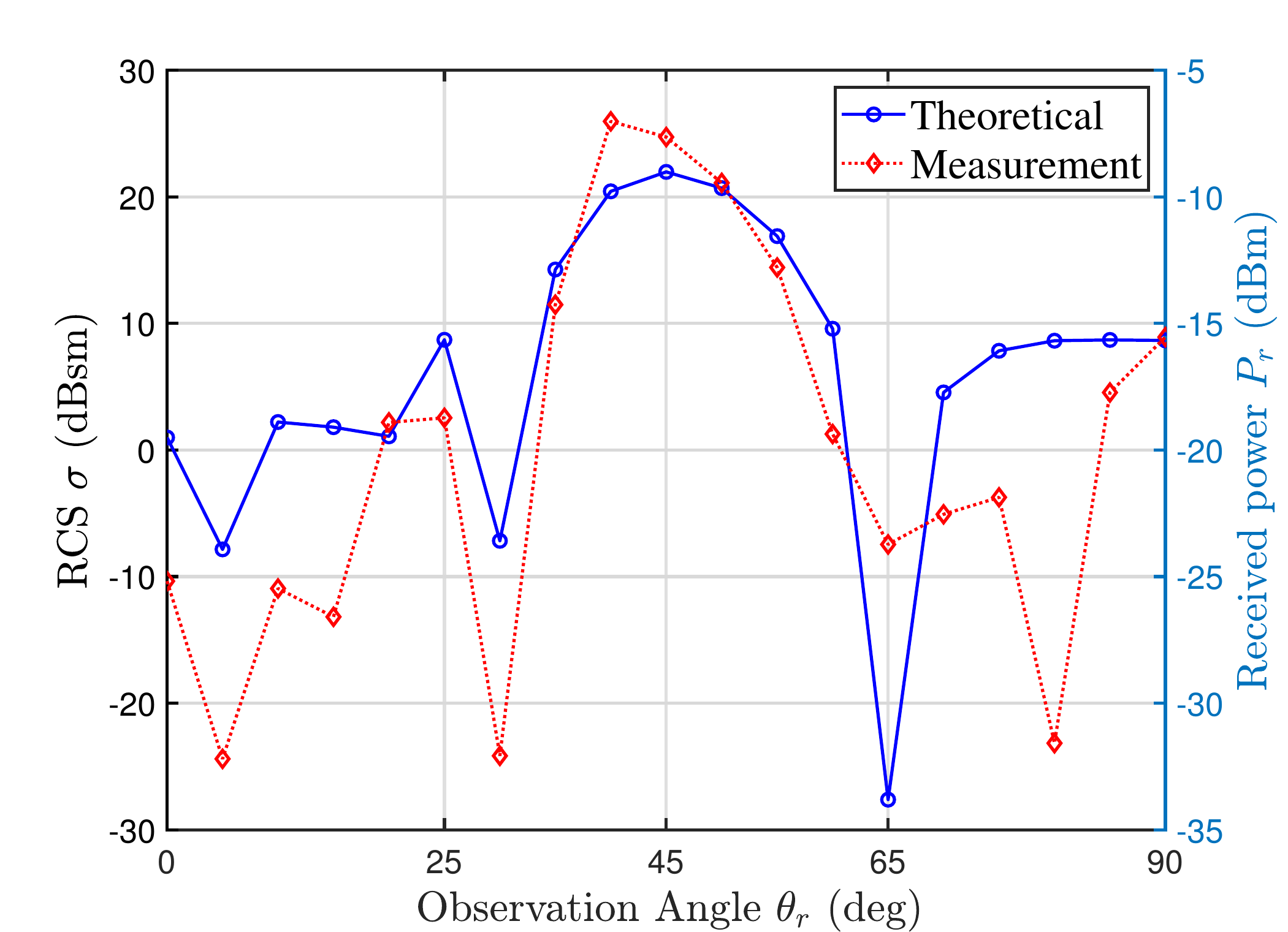}}
  \subfigure[$\theta_t=65^{\circ} $]{\includegraphics[width=1.98502in,height=1.5in]{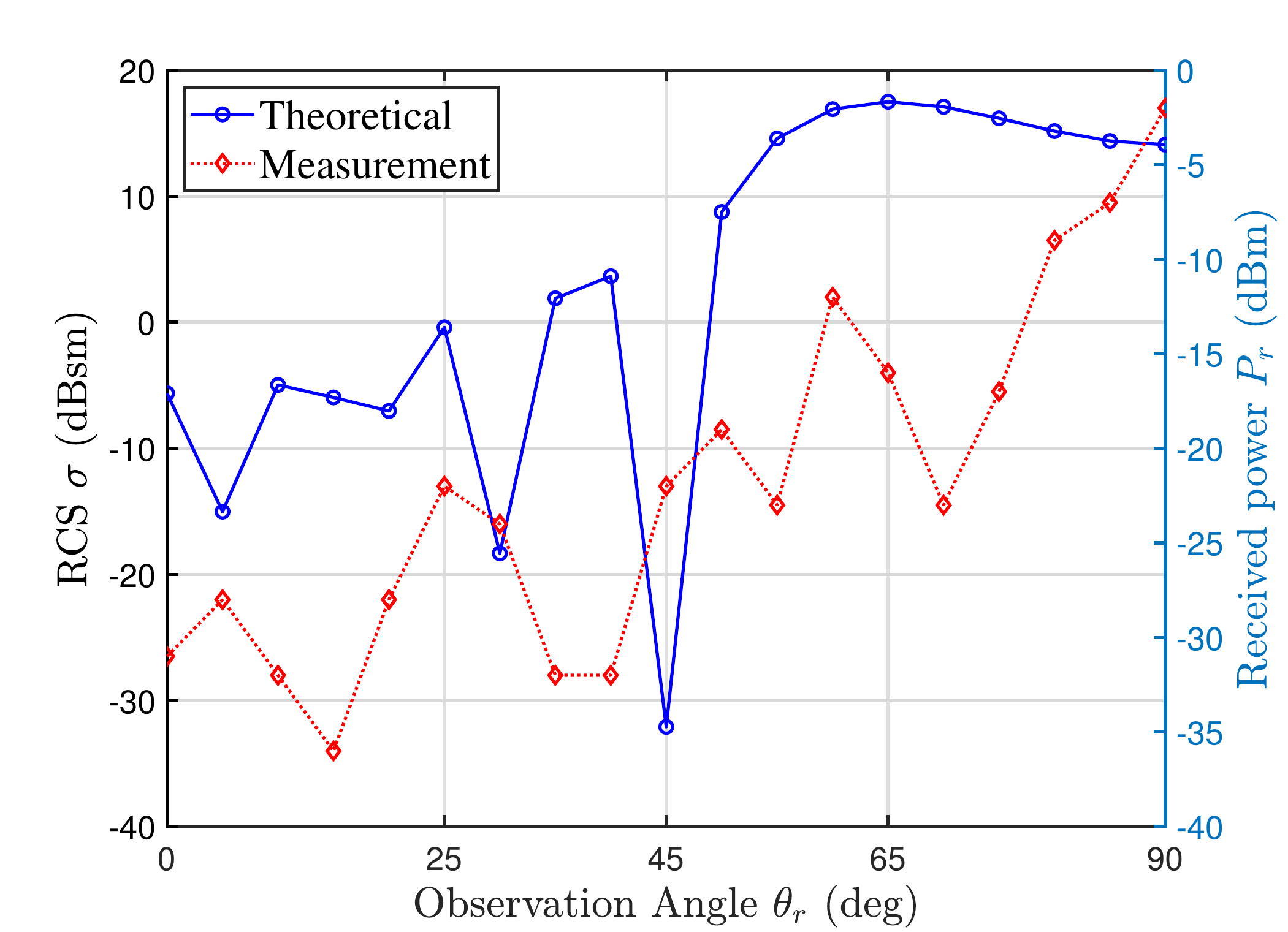}}
  \caption{Experiment and theoretical results for \eqref{case 1.1.1} with polarization angle $\varphi_t=90\,{\mathrm{^{\circ}}}$ or $270\,{\mathrm{^{\circ}}}$.}
  \label{experiment1}

\end{figure*}

\begin{figure*}[htbp]
  \centering
  \subfigure[$\theta_t=25^{\circ} $]{\includegraphics[width=1.98502in,height=1.5in]{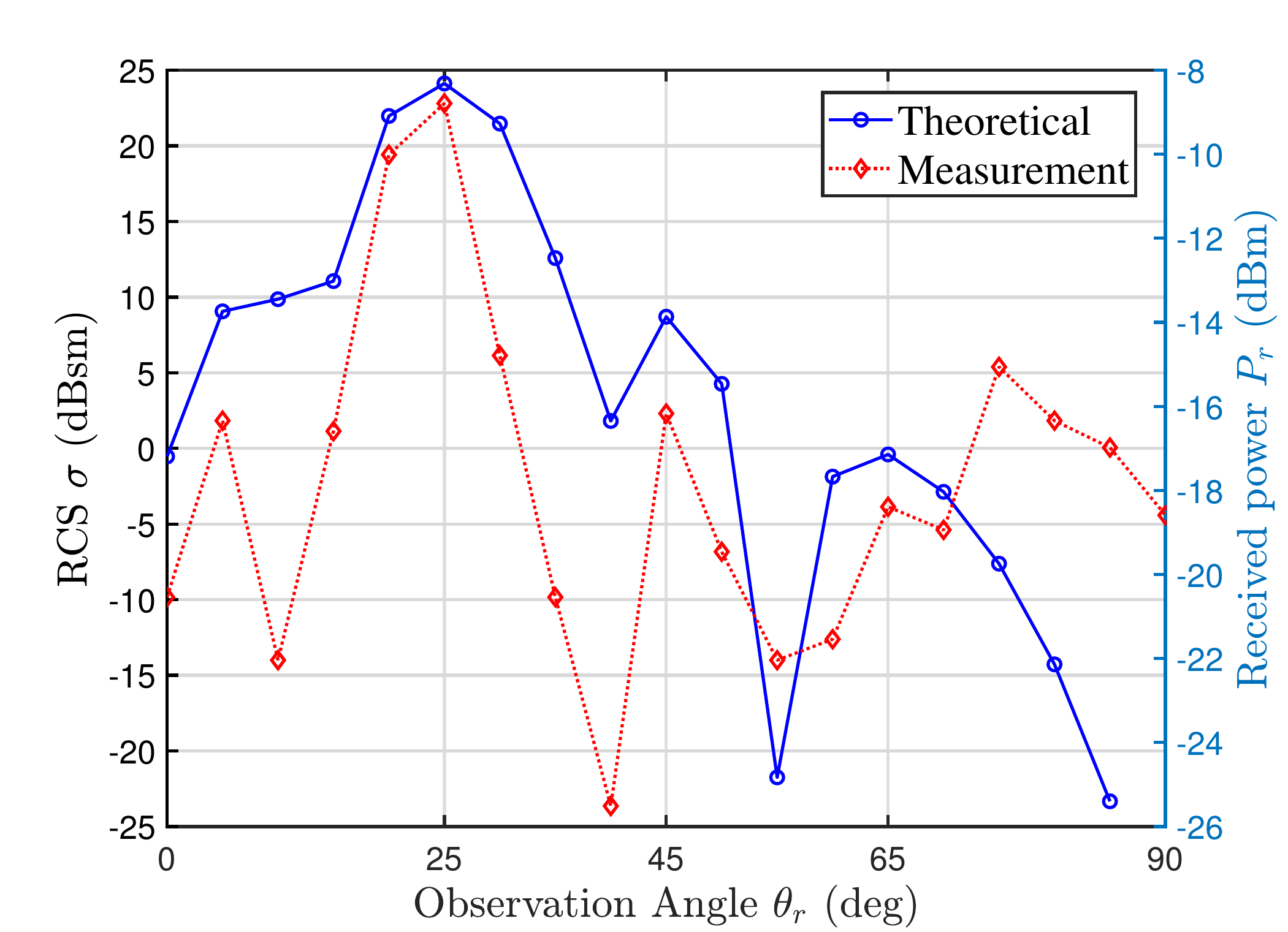}}
  \subfigure[$\theta_t=45^{\circ} $]{\includegraphics[width=1.98502in,height=1.5in]{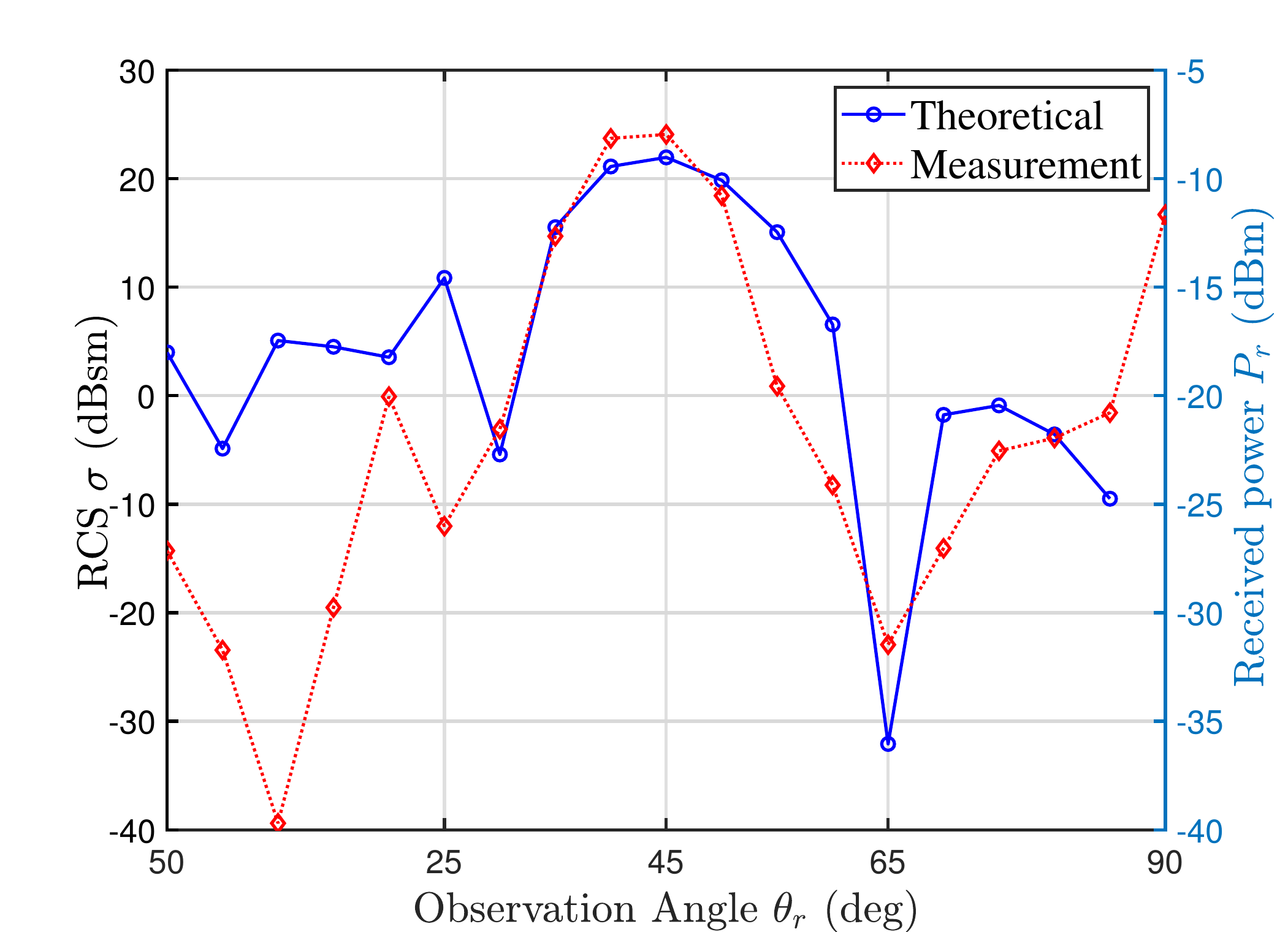}}
  \subfigure[$\theta_t=65^{\circ} $]{\includegraphics[width=1.98502in,height=1.5in]{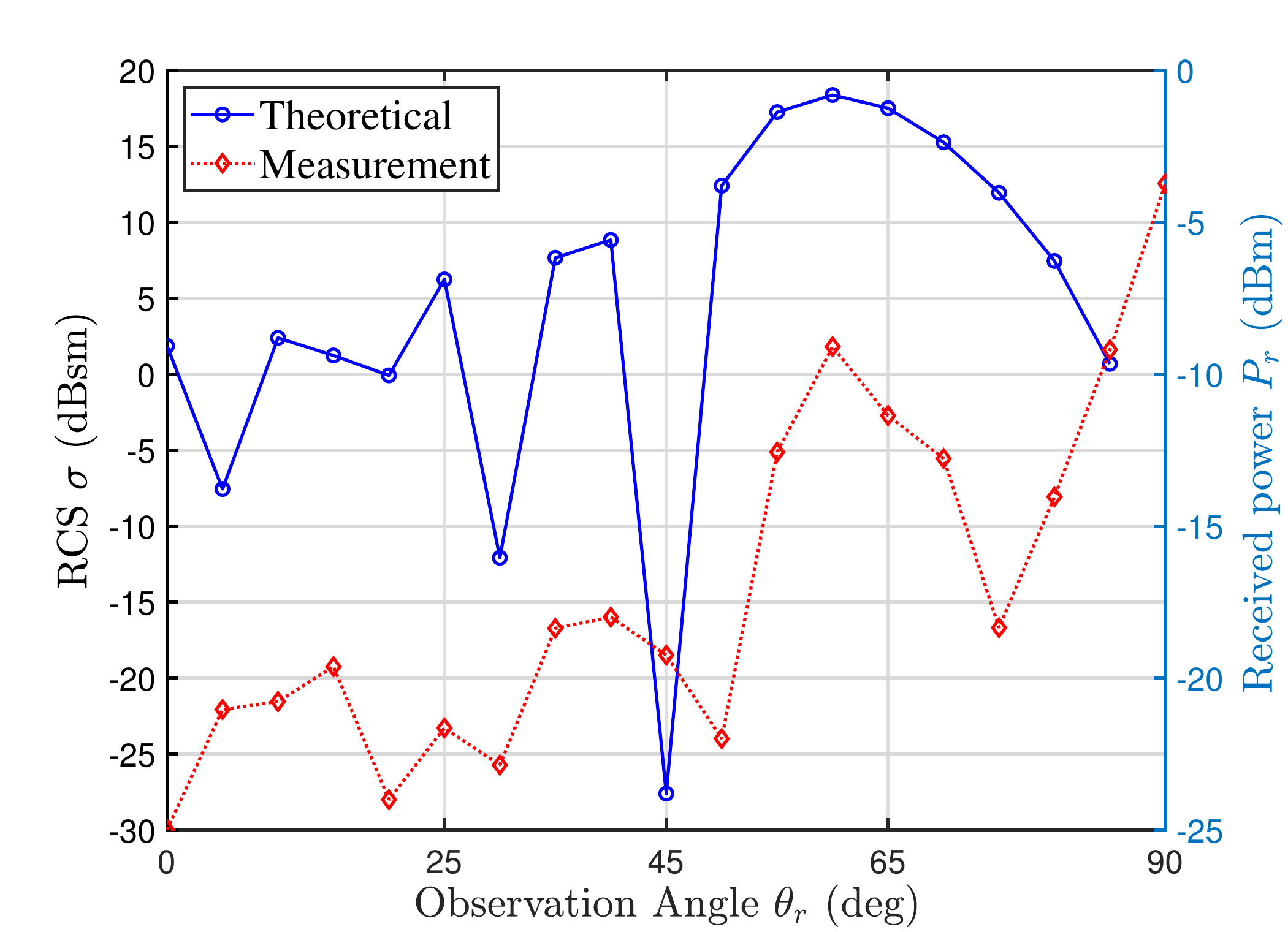}}
  \caption{Experiment and theoretical results for \eqref{case 1.2.1} with polarization angle $\varphi_t=0\,{\mathrm{^{\circ}}}$ or $180\,{\mathrm{^{\circ}}}$.}
  \label{experiment2}
  \vspace{-0.2cm}
\end{figure*}

\begin{table}[!t]
  \caption{MEASUREMENT PARAMETERS}
  \begin{center}
   \begin{tabular}{|c|c|}
  \hline
      $\text{Frequency}$ & $3\,{\mathrm{GHz}}$  \\ \hline
      $\text{Size of the metal plate}$ & $L_1=L_2=5\lambda$  \\ \hline
      $\text{Transmitting power}$ & $0\,{\mathrm{dBm}}$ \\ \hline
      $\text{Gain of the power amplifier}$ & $38.861\,{\mathrm{dB}}$ \\ \hline
      $\text{Gains of TX and RX antennas}$ & $16\,{\mathrm{dBi}}$ \\ \hline
      $\text{HPBW of antennas in $E$ and $H$ planes}$ & $33.31\,{\mathrm{^{\circ}}}$ and $30.81\,{\mathrm{^{\circ}}}$ \\ \hline
      $\text{Distance between RX and reflector}$ & $8\,{\mathrm{m}}$ \\ \hline     
      $\text{Distance between TX and reflector}$ & $8\,{\mathrm{m}}$ \\ \hline
      $\text{Incident angle }\theta_t$ & $25\,{\mathrm{^{\circ}}},45\,{\mathrm{^{\circ}}},65\,{\mathrm{^{\circ}}}$ \\ \hline 
      $\text{Observation angle }\theta_r$ & $0\,{\mathrm{^{\circ}}},5\,{\mathrm{^{\circ}}},...,85\,{\mathrm{^{\circ}}},90\,{\mathrm{^{\circ}}}$ \\ \hline 
      $\text{Polarization angle }\varphi_t$ & $0\,{\mathrm{^{\circ}}}$\text{ or }$180\,{\mathrm{^{\circ}}}\text{ and } 90\,{\mathrm{^{\circ}}}$\text{ or }$270\,{\mathrm{^{\circ}}}$ \\ \hline 
  \end{tabular}
  \label{MEASUREMENT PARAMETERS}

\end{center}
\vspace{-0.4cm}
\end{table}
Fig.~\ref{experiment1} and Fig.~\ref{experiment2} respectively plot the theoretically derived RCS in ${\mathrm{dBsm}}$ based on \eqref{case 1.1.1} and \eqref{case 1.2.1}, together with
the measured received power versus observation angles $\theta_r$ for three incident angles $\theta_t$,
where the polarization angles are $\varphi_t=90\,{\mathrm{^{\circ}}}$ or $270\,{\mathrm{^{\circ}}}$ for~\eqref{case 1.1.1} and $\varphi_t=0\,{\mathrm{^{\circ}}}$ or $180\,{\mathrm{^{\circ}}}$ for~\eqref{case 1.2.1}. 
It is firstly observed that for $\theta_t=25^\circ$ and $\theta_t=45^\circ$, 
the theoretical models match very well with the experimentally measured results.
In particular, the direction and beamwidth of the main lobe of the RCS agree quite well with that of
the measured power.
Furthermore, it is observed that similar to antenna arrays, the reflected waves by metal plates will propagate in the form of beams,
rather than along one single specular direction only.
This implies that metal reflectors, once properly deployed,  
could be used to enhance the coverage of dedicated areas, though it cannot achieve dynamic beam adjustment as by active antenna arrays or semi-passive IRS.  
Furthermore, it is observed that the side lobe level of metal reflectors can be significantly lower than that of the
main lobe, e.g., by more than $10\,{\mathrm{dB}}$, and the gap is expected to be even higher if larger metal plate is used.
Another observation is that for $\theta_t=65^\circ$, though the theoretical RCS model reflects the general trend of the
measured results, the match is not as good as the other two cases.
This is probably caused by the interference of the direct link from the transmitting to receiving antenna, which becomes more severe when the incident angle is large
due to the signal leakage from the mainlobe and sidelobe of horn antennas.

\section{Conclusions}
In this paper, we derived a general reflection model that is applicable to metal plates of any size, any orientation, and any linear polarization. 
Different from existing models, the derived model is given compactly in terms of the RCS of the metal plate, as a function of its physical dimensions and
orientation vectors, as well as the wave polarization and the wave deflection vector. 
Furthermore, experimental results based on actual field measurements are provided to validate the accuracy of our developed model.

\vspace{-0.2cm}
\section*{Acknowledgment}
This work was supported by the National Key R\&D Program of China with Grant number 2019YFB1803400.
\vspace{-0.2cm}
\begin{appendices}
  \section{Proof of theorem~\ref{RCS}}
  By substituting~\eqref{js} into~\eqref{etheta} and~\eqref{ephi}, we have
\vspace{-0.2cm}
\begin{equation}
  E^{r}_{\theta} = -\frac{j k \eta H_0 e^{-j k d_r}}{2 \pi d_r}    ({\bf{n}}\times{\bf{a}}_H)\cdot{\bf{a}}_{r\theta}    \iint_{S} e^{jk \left({\bf{a}}_r-{\bf{a}}_t \right) \cdot {\bf{r}}^{\prime}} {\rm d} s,
\end{equation}
\begin{equation}
  E^{r}_{\phi} = -\frac{j k \eta H_0 e^{-j k d_r}}{2 \pi d_r}  \!  ({\bf{n}}\times{\bf{a}}_H)\cdot{\bf{a}}_{r\phi}   \!      \iint_{S} e^{jk \left({\bf{a}}_r-{\bf{a}}_t \right) \cdot {\bf{r}}^{\prime}} {\rm d} s.
\end{equation}

  To evaluate the integral $\iint_{S} e^{jk \left({\bf{a}}_r-{\bf{a}}_t \right) \cdot {\bf{r}}^{\prime}} {\rm d} s$,
  we express the point ${\bf{r}}^{\prime}$ on the metal plate as ${\bf{r}}^{\prime}=\alpha{\bf{l}}_1+\beta{\bf{l}}_2$ for some $\alpha$ and $\beta$. Thus,
  \begin{align}
    &\iint_{S} e^{jk \left({\bf{a}}_r-{\bf{a}}_t \right) \cdot {\bf{r}}^{\prime}} {\rm d} s\nonumber\\
    &=\int_{-L_2/2}^{L_2/2}\int_{-L_1/2}^{L_1/2} e^{jk \left({\bf{a}}_r-{\bf{a}}_t \right) \cdot ({\alpha{\bf{l}}_1+\beta{\bf{l}}_2})} {\rm d} \alpha{\rm d} \beta\nonumber\\
    &=\int_{-L_2/2}^{L_2/2} e^{jk \beta \left({\bf{a}}_r-{\bf{a}}_t \right) \cdot {{\bf{l}}_2}}   \int_{-L_1/2}^{L_1/2} e^{jk \alpha \left({\bf{a}}_r-{\bf{a}}_t \right) \cdot {{\bf{l}}_1}} {\rm d} \alpha{\rm d} \beta \nonumber\\
    &=L_1 L_2 \sinc \!\left( \frac{k L_1}{2} \left({\bf{a}}_r-{\bf{a}}_t \right) \!\cdot\! {\bf{l}}_1 \right)  \sinc \!\left( \frac{k L_2}{2} \left({\bf{a}}_r-{\bf{a}}_t \right) \!\cdot\! {\bf{l}}_2 \right),
  \end{align}
where $\sinc(x)=\sin(x)/{x}$.
The squared magnitude of the reflected electric field can be expressed as
\vspace{-0.1cm}
\begin{align}
    \left|{\bf{E}}^{r}\right|^{2}&=\left|E^r_{\theta}\right|^{2}+\left|E^r_{\phi}\right|^{2} \nonumber\\
    &=\frac{\eta^2 H_{0}^2 L_{1}^2 L_{2}^2}{\lambda^2 d_r^2} \left( \left[({\bf{n}}\times{\bf{a}}_H)\cdot{\bf{a}}_{r\theta} \right]^2+ \left[({\bf{n}}\times{\bf{a}}_H)\cdot{\bf{a}}_{r\phi}  \right]^2 \right) \nonumber \\
    &\times \sinc^2 \!\left( \frac{k L_1}{2} \left({\bf{a}}_r-{\bf{a}}_t \right) \!\cdot \!{\bf{l}}_1 \right)   \sinc^2 \!\left( \frac{k L_2}{2} \left({\bf{a}}_r-{\bf{a}}_t \right) \!\cdot \!{\bf{l}}_2 \right).
\end{align}

The squared magnitude of the incident electric field is $|{\bf{E}}^t|^2=\eta ^2 H_0^2$, and the RCS of the metal plate can be obtained as
\vspace{-0.1cm}
\begin{align}
  \sigma &=\lim _{d_r \rightarrow \infty} 4 \pi d_r^{2} \frac{\left|{\bf{E}}^{r}\right|^{2}}{\left|{\bf{E}}^{t}\right|^{2}}\\
  &=\frac{4\pi L_1^2 L_2^2}{\lambda^2}\left( \left[({\bf{n}}\times{\bf{a}}_H)\cdot{\bf{a}}_{r\theta} \right]^2+ \left[({\bf{n}}\times{\bf{a}}_H)\cdot{\bf{a}}_{r\phi}  \right]^2 \right)  \nonumber\\
  &\times \sinc^2 \!\left( \frac{k L_1}{2} \left({\bf{a}}_r-{\bf{a}}_t \right) \!\cdot \!{\bf{l}}_1 \right)   \sinc^2 \!\left( \frac{k L_2}{2} \left({\bf{a}}_r-{\bf{a}}_t \right) \!\cdot\! {\bf{l}}_2 \right)\nonumber\\
  &=\frac{4\pi L_1^2 L_2^2}{\lambda^2}|| \left( {\bf{n}} \times {\bf{a}}_H \right) \times {\bf{a}}_r||^2\nonumber\\
  &\times \sinc^2 \!\left( \frac{k L_1}{2} \left({\bf{a}}_r-{\bf{a}}_t \right) \!\cdot \!{\bf{l}}_1 \right)   \sinc^2 \!\left( \frac{k L_2}{2} \left({\bf{a}}_r-{\bf{a}}_t \right) \!\cdot\! {\bf{l}}_2 \right).
\end{align}

\vspace{0.0cm}

  \section{Proof of~\eqref{defineat}-\eqref{defineah}}
  \begin{figure}[!t]
    \setlength{\abovecaptionskip}{-0.1cm}
    \setlength{\belowcaptionskip}{-0.3cm}
    \centering
    \centerline{\includegraphics[scale=0.19]{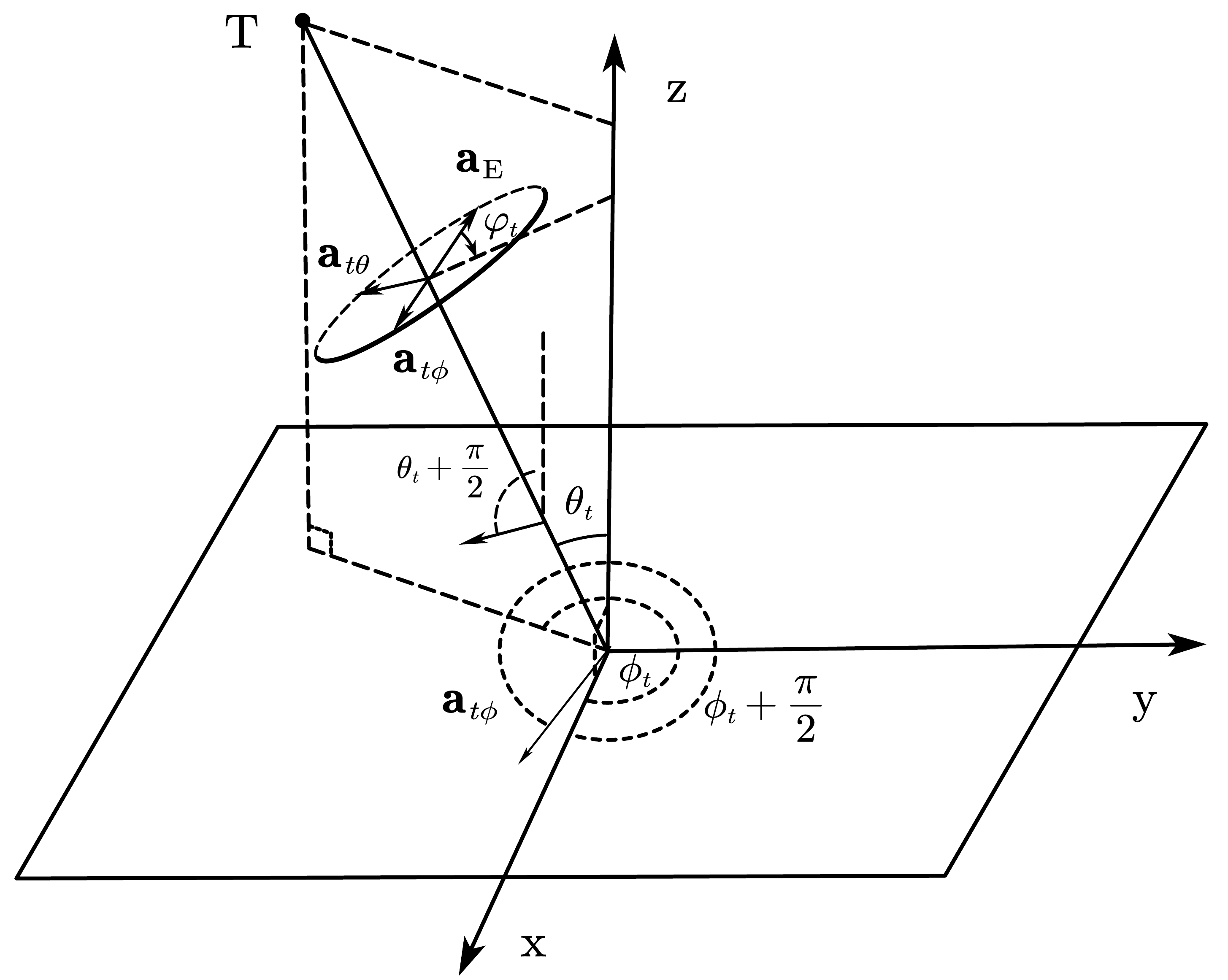}}
    \caption{Illustration for angles in~\eqref{defineah}.}
    \label{modelofappe}
    \vspace{-0.0cm}
  \end{figure}
  
For a unit vector with zenith angle $\theta$ and azimuth angle $\phi$, it can be expressed as $(1,\theta,\phi)$ in spherical coordinate system
and $\left(\sin \theta \cos \phi,\sin \theta\sin \phi,\cos \theta\right)$ in Cartesian coordinate system.
Thus, ${\bf{a}}_{r}$ and ${\bf{a}}_{t}$ can be given by

\begin{equation}
  {\bf{a}}_{r}=\left(\sin \theta_{r} \cos \phi_{r},\sin \theta_{r} \sin \phi_{r},\cos \theta_{r}\right),
\end{equation}
\begin{equation}
  {\bf{a}}_{t}=\left(-\sin \theta_{t} \cos \phi_{t},-\sin \theta_{t} \sin \phi_{t},-\cos \theta_{t}\right).
\end{equation}

As illustrated in Fig. \ref{modelofappe}, ${\bf{a}}_{E}$ can be expressed as 
\begin{equation}
  {\bf{a}}_{E}=-\cos\phi_t {\bf{a}}_{t\theta}-\sin\varphi_t{\bf{a}}_{t\phi},
\end{equation}
where ${\bf{a}}_{t\theta}$ lies in the plane formed by $z$-axis and the line $OT$ and is perpendicular to the line $OT$.
Therefore, ${\bf{a}}_{t\theta}$ can be expressed as $(1,\theta_t+\frac{\pi}{2},\phi_t)$ in spherical coordinate system, i.e.,
\begin{equation}
  {\bf{a}}_{t\theta}=\left(\cos \theta_{t} \cos \phi_{t}, \cos \theta_{t} \sin \phi_{t},-\sin \theta_{t}\right).
\end{equation}

Similarly, ${\bf{a}}_{t\phi}$ is perpendicular to the plane formed by $z$-axis and the line $OT$.
Therefore, ${\bf{a}}_{t\phi}$ can be expressed as $(1,\frac{\pi}{2},\phi_t+\frac{\pi}{2})$ in spherical coordinate system, i.e., 
\begin{equation}
  {\bf{a}}_{t\phi}=\left(-\sin \phi_{t}, \cos \phi_{t}, 0\right).
\end{equation}

Therefore, ${\bf{a}}_{H}$ can be easily obtained by ${\bf{a}}_{t}\times{\bf{a}}_{E}$.

\end{appendices}

\vspace{0.6cm}
\bibliographystyle{IEEEtran}
\bibliography{ref1}

\end{document}